\documentclass[a4paper,12pt]{article}
\pdfoutput=1

\usepackage{color}
\definecolor{rosso}{cmyk}{0,1,1,0.4}
\definecolor{rossos}{cmyk}{0,1,1,0.55}
\definecolor{rossoc}{cmyk}{0,1,1,0.2}
\definecolor{blu}{cmyk}{1,1,0,0.3}
\definecolor{blus}{cmyk}{1,1,0,0.6}

\usepackage[dvipsnames]{xcolor} 
\usepackage{lscape}
\usepackage{colortbl}
\usepackage{multirow} % merge rows in table
\usepackage{tabularx}
\usepackage{pbox}
\usepackage{hhline}
\usepackage{epstopdf}

\usepackage{url}
\usepackage{amsmath}
\usepackage[final]{graphicx}% Include figure files
\usepackage{caption}
\usepackage{subcaption}%subfigures
\usepackage{bm}% bold math
\usepackage{amssymb}

\begin{document}
\hfill
\begin{minipage}{20ex}\small
ZTF-EP-16-02\\
\end{minipage}

\begin{center}
{\LARGE \bf 
Radiative neutrino models\\  in light of diphoton signals\\}
\vspace{0.3in}
{\bf 
Oleg Antipin$^{a}$, 
Petar~\v{C}uljak$^{b}$,
Kre\v{s}imir~Kumeri\v{c}ki$^{b}$  
and Ivica~Picek$^{b}$
}\\[4ex]
%\begin{flushleft}
{\sl
$^{a}$Rudjer Bo\v{s}kovi\'c Institute, Division of Theoretical Physics,
Bijeni\v{c}ka 54, HR-10000 Zagreb, Croatia\\[1.5ex]
$^{b}$Department of Physics, Faculty of Science, University of Zagreb,
 P.O.B. 331, HR-10002 Zagreb, Croatia\\}
%\end{flushleft}

%\vspace{0.5in}
%\today \\[5ex]
\end{center}

\vspace{0.2in}
\begin{center}\large\bf Abstract\end{center}

Viable explanations of a hinted 750 GeV scalar resonance may be sought
within the extensions of the SM Higgs sector aimed at generating neutrino masses at the loop level.
We confront a compatibility with the 750 GeV diphoton excess for two recent models which do not 
need to impose ad hoc symmetry to forbid the tree-level masses: a one-loop mass model providing
the $H(750)$ candidate within its real triplet scalar representation and a three-loop mass model providing it within its two Higgs doublets. Besides accounting for the 750 GeV resonance, we demonstrate that these complementary neutrino-mass scenarios
have different testable predictions for the LHC which should show up soon as more data is accumulated during the ongoing 13 TeV run.

\vspace{0.2in}

%\begin{flushleft}
%\small
%\emph{PACS}:12.60.Fr; 14.60.Pq
%\\
%\emph{Keywords}: Extensions of Higgs sector; Neutrino mass; Diphoton excess
%\end{flushleft}

\clearpage
%\tableofcontents

\section{Introduction}

After discovery of the $125$ GeV Standard Model (SM) Higgs boson~\cite{Aad:2012tfa,Chatrchyan:2012ufa}, there are alluring hints 
of a new scalar resonance responsible for the diphoton excess at $750$ GeV in the ongoing run of the LHC~\cite{Aaboud:2016tru,Khachatryan:2016hje}. \\
Most of the existing studies which interpret the hinted resonance as an indication of a second Higgs boson, consider it in framework with an additional scalar singlet or with a second scalar doublet. In both cases one maintains the value of the electroweak precision parameter $\rho = 1$ at the tree level; while for the scalar singlet this is obvious this issue has been studied in detail for two Higgs doublet  models (2HDM) in~\cite{Branco:2011iw}. Thereby it was found that 2HDM cannot accommodate recent diphoton excess without introducing 
additional massive particles~\cite{Angelescu:2015uiz,Altmannshofer:2015xfo,Bertuzzo:2016fmv,Murphy:2015kag,DiChiara:2015vdm}. If we employ instead the scalar
field in a weak triplet representation, it is still possible to keep the $\rho$-parameter protected by using both real and complex triplet scalar fields, like in the custodial triplet model known as the Georgi-Machacek model~\cite{Georgi:1985nv}. It has been introduced as another benchmark model for a diphoton study in~\cite{Fabbrichesi:2016alj} and~\cite{Chiang:2016ydx}.\\
We study a possible appearance of the hinted resonance in the context of beyond-SM (BSM) fields which appear in models of radiative neutrino masses. 
Specifically, we confront the capacity to fit the $750$ GeV excess of two different radiative neutrino mass scenarios displayed in Table~\ref{models}: 
\begin{enumerate}
\item The one-loop neutrino mass model~\cite{Brdar:2013iea} with minimal BSM representations 
providing the neutral component of a real scalar field $\Delta$  in the adjoint representation of the $SU(2)_L$ as the 750 GeV resonance candidate. 
\item The three-loop neutrino mass model~\cite{Culjak:2015qja} with exotic BSM representations
where the 750 GeV candidate emerges in the form of
the heavy CP-even neutral scalar field in the framework of the 2HDM.
\end{enumerate}

\begin{table}[ht]
\footnotesize
\begin{center}
\begin{tabularx}{\textwidth}{| *{4}{>{\centering\arraybackslash} c|} | *{4}{>{\centering\arraybackslash} c|}}
 \hhline{----||----}
 \multicolumn{4}{|c||}{\raisebox{-0.5ex}{\includegraphics[width=0.45\textwidth]{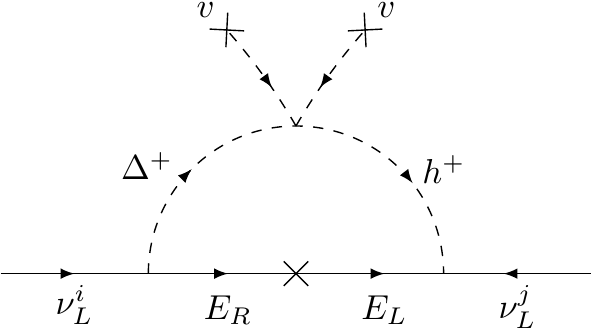}}} & \multicolumn{4}{c|}{\raisebox{-0.5ex}{\includegraphics[width=0.45\textwidth]{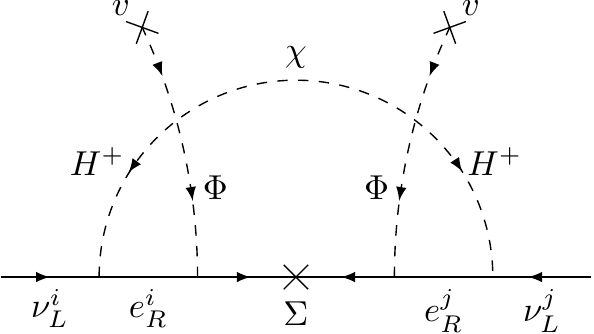}}} \\
 \hhline{----||----}
 \rowcolor[cmyk]{0.2,0.0,0,0.0}  Name &  $SU(2)_L$ &  $U(1)_Y$ & $Q$ &  Name &  $SU(2)_L$ &  $U(1)_Y$ &  $Q$ \\
 \hhline{----||----}
 \rowcolor[cmyk]{0,0,0.2,0} $\textcolor{LimeGreen}{\boldsymbol{\Delta}}$ & $3$ & $0$ & $\pm1,0$ & $\textcolor{LimeGreen}{\boldsymbol{H}}_{\textcolor{LimeGreen}{\boldsymbol{1,2}}}$ & $2$ & $1$  & $0,1$ \\
 \hhline{----||----}
  \rowcolor[cmyk]{0,0,0.2,0} $h^+$ & $1$ & $2$  & $1$ & $ \Phi$ & $5$  & $-2$ &$-3,-2,\pm 1,0$ \\
 \hhline{----||----}
 \rowcolor[cmyk]{0,0.2,0,0.1}$E_{R}$ & $2$ & $-1$ & $0,-1$ & \cellcolor[cmyk]{0,0,0.2,0} $\chi$ & \cellcolor[cmyk]{0,0,0.2,0} $7$ & \cellcolor[cmyk]{0,0,0.2,0} $0$  & \cellcolor[cmyk]{0,0,0.2,0} $\pm3,\pm2,\pm1,0$\\
 \hhline{----||----}
 \rowcolor[cmyk]{0,0.2,0,0.1}$E_{L}$ & $2$ & $-1$  & $0,-1$ & $\Sigma $ & $5$ & $0$ & $\pm2,\pm1,0$\\
 \hhline{----||----}
\end{tabularx}
\end{center}
\caption{ \rm  Neutrino mass models. Scalar fields are in (light) yellow and fermion fields in three generations are in (dark) red. The fields containing the 750 GeV candidate are in (light grey) green. For the one-loop model (left) the SM Higgs doublet manifests itself only via its VEV $v$ in the neutrino mass diagram. 
% * <antipin@cp3-origins.net> 2016-06-14T13:38:25.139Z:
%
% ^.
}
\label{models}
\end{table}

The paper is structured as follows. In Sec.~\ref{Review} we briefly review these radiative neutrino mass models and study their implications for the diphoton signal in Sec.~\ref{diphoton}. 
We discuss the stability of the scalar potential as well as Landau poles of relevant couplings in Sec.~\ref{perturbativity}
and present our conclusions in Sec.~\ref{Conclusions}.

\section{Two radiative neutrino mass models} \label{Review}

\subsection{The one-loop model}

The first mass model~\cite{Brdar:2013iea} in our focus is based on the one-loop diagram displayed on the LHS in Table~\ref{models}. It 
has an appeal to invoke low non-singlet weak representations and to be free of imposing an additional ad hoc $Z_2$ symmetry to eliminate the tree-level contribution. Still,  a Dark Matter (DM) stabilizing $Z_2$ symmetry is needed in the proposed attempts to account for the DM in ``inert triplet'' variants: the one realized  with a $Z_2$ odd real triplet in~\cite{Brdar:2013iea,Law:2013dya,Law:2013saa} or another with a $Z_2$ odd complex scalar triplet~\cite{Okada:2015vwh}. However, we will not consider here such cases where the new scalar field doesn't mix with the SM Higgs field.\\
Our model may be viewed as a substitute for the original 
one-loop neutrino-mass model by Zee~\cite{Zee:1980ai} which, in meantime, has been ruled out by data: a charged scalar singlet $h^+ \sim (1,2)$ in Zee loop-diagram has been kept,
while its second Higgs doublet has been replaced by  
hypercharge zero triplet scalar field
\begin{equation}
\Delta=\frac{1}{\sqrt{2}}\sum_{j}\sigma_{j}\Delta^{j}=
    \left(  \begin{array}{ccc}
       \frac{1}{\sqrt{2}} \Delta^0 & \Delta^+\\
       \Delta^- & -\frac{1}{\sqrt{2}} \Delta^0
    \end{array} \right) \sim (3,0)\ ,
\end{equation}
in conjunction with the vector-like lepton $E_{R,L} \sim (2,-1)$ in three generations.
Such modification of Zee model may be interesting in light of some findings that possible explanation of 750 GeV resonance requires both scalar and fermion BSM fields.
The gauge invariant scalar potential of this model reads
\begin{eqnarray}\label{potential}
&&V(H,\Delta,h^+)= -\mu_H^2 H^\dag H + \lambda_1(H^\dag H)^2 + \mu_h^2 h^- h^+ + \lambda_2 (h^- h^+)^2\nonumber\\
&&+ \mu_\Delta^2 \mathrm{Tr}[{\Delta}^2] + \lambda_3 (\mathrm{Tr}[\Delta^2])^2 + \lambda_4 H^\dag H h^-h^+ + \lambda_5 H^\dag H \mathrm{Tr}[\Delta^2] \nonumber\\
&&+ \lambda_6 h^-h^+ \mathrm{Tr}[\Delta^2] + (\lambda_7 H^\dag\Delta\tilde{H}h^+ + \mathrm{H.c.}) + \mu H^\dag \Delta H \ ,
\end{eqnarray}
where the vacuum expectation value (VEV) $v=246$ GeV 
of the neutral component of the Higgs doublet $H=(\phi^+,\phi^0)^T$  leads to electroweak symmetry breaking (EWSB).
Without imposing $Z_2$ symmetry the trilinear $\mu$ term 
in (\ref{potential}) leads to an induced VEV  $\langle \Delta^0\rangle$ for the neutral triplet component, which is constrained by electroweak measurements to be smaller than a few GeV.\\

\underline{\bf Neutrino mass:} The effective neutrino mass operator is generated via the $\lambda_7$ coupling in (\ref{potential}) and appropriate  Yukawa interactions from a gauge invariant Lagrangian
\begin{eqnarray}\nonumber
\mathcal{L}&=& M \overline{E_L} E_R + y \overline{E_L} H l_R + g_1 \overline{(L_L)^c} E_L h^+\\
&+& g_2 \overline{L_L} \Delta E_R + g_3 \overline{E_L} \Delta E_R + g_4 \overline{(L_L)^c} L_L h^+ + \mathrm{h.c.} \ .
\label{Yukawa1loop}
\end{eqnarray}
Here $y$ and $g_{1,2,3,4}$ are the Yukawa-coupling matrices and for simplicity we drop the flavour indices altogether.
The resulting neutrino mass reads~\cite{Brdar:2013iea}
\begin{eqnarray}
\label{effective}
\mathcal{M}_{ij}&=&\sum_{k=1}^3\frac{[(g_1)_{ik} (g_2)_{jk} + (g_2)_{ik}(g_1)_{jk}]} {16\pi^{2}} \ \lambda_7 \; v^2 \; M_{E_k} \\
\hspace{1.8cm}
&&\frac{M_{E_k}^{2}m_{h^+}^{2}\ln{\frac{M_{E_k}^{2}}{m_{h^+}^{2}}}+
M_{E_k}^{2}m_{\Delta^+}^{2}\ln{\frac{m_{\Delta^+}^{2}}{M_{E_k}^{2}}}
+m_{h^+}^{2}m_{\Delta^+}^{2}\ln{\frac{m_{h^+}^{2}}{m_{\Delta^+}^{2}}}}{({m_{h^+}^{2}-m_{\Delta^+}^{2}})
(M_{E_k}^{2}-m_{h^+}^{2})(M_{E_k}^{2}-m_{\Delta^+}^{2})} \nonumber\; .
\end{eqnarray}
Assuming
%as in~\cite{Brdar:2013iea}, 
the mass values in the diphoton-preferred range, as we will use later, $M_E \sim m_{\Delta^+} \sim m_{h^+} \sim 400$ GeV, (\ref{effective}) leads to $m_\nu \sim 0.1$ eV
for the couplings $g_{1,2}$ and $\lambda_7$ of the order of $10^{-4}$.

\subsection{The three-loop model}

The second mass model~\cite{Culjak:2015qja} in our focus is based on the three-loop diagram displayed on the RHS in Table~\ref{models}.
%Fig.~\ref{a-b-numass}(b).
Notably, this model contains 2HDM sector augmented by exotic scalar multiplets needed to close the three-loop mass diagram and motivated by the minimal dark matter (MDM) setup~\cite{Cirelli:2005uq}: the complex scalar pentuplet $\Phi$ and a real scalar field $\chi$ in the septuplet representation. Since $\Phi$ and $\chi$ fields do not form gauge invariant couplings with the SM particles, there is again no need for an additional symmetry to eliminate the tree-level neutrino mass contributions. This model is ideally suited for producing small neutrino masses with non-suppressed couplings and the multiply-charged components in similar setup have already been claimed to explain the 750 GeV diphoton excess~\cite{Moretti:2015pbj,Han:2015qqj}. 

The three-loop model at hand is in addition fortuitously scotogenic~\cite{Culjak:2015qja}:
a standard discrete $\tilde Z_2$ symmetry imposed to produce a natural flavour conservation in 2HDM results in accidental $Z_2$ odd parity of its BSM sector shown in Table~\ref{Z-charges}.
On account of it, the lightest among the three generations  ($\alpha =1,2,3$) of exotic real fermions $\Sigma_{\alpha} \sim (5,0)$ turns out to be a viable DM candidate. 
Out of four different ways the Higgs doublets are 
conventionally assigned charges under a $\tilde Z_2$ symmetry~\cite{Kanemura:2014bqa}, we adopt 
the  ``lepton-specific" (Type X or Type IV) 2HDM implemented originally in~\cite{Aoki:2008av, Aoki:2009vf} and shown in Table~\ref{Z-charges} .
\begin{table}
%\small
\begin{center}
  \begin{tabular}{|c|ccccc|cc|ccc|}
   \hline
 \hbox{Symmetry} & $Q_i$ & $u_{i R}$ & $d_{i R}$ & $L_{i L}$ & $e_{i R}$ & $H_{\bf{1}}$ & $H_{\bf{2}}$ & $\Phi$ &
    $\chi$ & $\Sigma_{\alpha}$ \\\hline
\rowcolor[cmyk]{0.1,0,0.1,0}
\cellcolor[cmyk]{0.0,0,0.2,0}$Z_2\frac{}{}$                {\rm accidental} & $+$ & $+$ & $+$ & $+$ & $+$ & $+$ & $+$ & $-$ & $-$ & $-$ \\ \hline  
 \rowcolor[cmyk]{0.2,0.0,0,0.0}
  \cellcolor[cmyk]{0,0.2,0.0,0}
  $\tilde{Z}_2\frac{}{}$ {\rm imposed}& $+$ & $-$ & $-$ & $+$ &
                       $+$ & $+$ & $-$ & $+$ & $-$ & $+$ \\\hline
   \end{tabular}
\end{center}
  \caption{Charge assignment under an automatic $Z_2$ symmetry which is induced by the imposed $\tilde Z_2$ symmetry 
  in the lepton-specific 2HDM.}
  \label{Z-charges}
\end{table}
In terms of physical fields, the two Higgs doublet fields $H_{\bf{1,2}}\sim (2,1)$  are written as
\begin{equation}
H_{\bf{1}}=\left(\begin{array}{c}
\displaystyle G^+\cos\beta -H^+\sin\beta  \\
\displaystyle \frac{1}{\sqrt{2}}\left(v_1-h\sin\alpha+H\cos\alpha+\mathrm{i}\left( G\cos\beta-A\sin\beta \right)\right)
\end{array}
\right),
\end{equation}
\begin{equation}
H_{\bf{2}}=\left(\begin{array}{c}
\displaystyle G^+\sin\beta +H^+\cos\beta  \\
\displaystyle \frac{1}{\sqrt{2}}\left(v_2+h\cos\alpha+H\sin\alpha+\mathrm{i}\left( G\sin\beta+A\cos\beta \right)\right)
\end{array}
\right),
\end{equation}
and their electroweak VEVs define $\tan\beta\equiv v_2 /v_1$.
The physical charged scalars are $H^\pm$, and, besides the three Goldstone bosons ($G, G^\pm$)
eaten by $Z$ and $W^\pm$, there is a CP-odd physical neutral scalar $A$.
The two CP-even neutral states  $h$ and $H$ (mixing with the angle $\alpha$) are proposed to be the physical Higgs fields $h(125)$ and $H(750)$.

Conventionally, the VEVs $v_1=v\cos\beta$ and $v_2=v\sin\beta$ (which are related to the SM VEV $v=$ 246 GeV by $v^2 = v_1^2 + v_2^2$) originate from  $m_{11}^2$ and $m_{22}^2$ terms 
through the minimization conditions of the most general CP-conserving 2HDM potential 
\begin{equation}
\begin{split}
V(H_{\bf{1}},H_{\bf{2}})  &= m_{11}^2 H_{\bf{1}}^\dagger H_{\bf{1}}+ m_{22}^2 H_{\bf{2}}^\dagger H_{\bf{2}}
-[m_{12}^2 H_{\bf{1}}^\dagger H_{\bf{2}}+ \, \text{h.c.} ] \\
& +\frac{1}{2}\lambda_1(H_{\bf{1}}^\dagger H_{\bf{1}})^2
+\frac{1}{2}\lambda_2(H_{\bf{2}}^\dagger H_{\bf{2}})^2\\
& +\lambda_3(H_{\bf{1}}^\dagger H_{\bf{1}})(H_{\bf{2}}^\dagger H_{\bf{2}})
+\lambda_4(H_{\bf{1}}^\dagger H_{\bf{2}})(H_{\bf{2}}^\dagger H_{\bf{1}}) \\
& +\left\{\frac{1}{2}\lambda_5(H_{\bf{1}}^\dagger H_{\bf{2}})^2
+\big[\lambda_6(H_{\bf{1}}^\dagger H_{\bf{1}})
+\lambda_7(H_{\bf{2}}^\dagger H_{\bf{2}})\big]
H_{\bf{1}}^\dagger H_{\bf{2}}+\, \text{h.c.}\right\}\,.
\label{2HDpot}
\end{split}
\end{equation}
It is possible to trade the five quartic couplings $\lambda_1$ to $\lambda_5$ for the four physical Higgs boson masses (as free input parameters) and the mixing parameter $\sin(\beta-\alpha)$.

The additional exotic scalar fields $\Phi \sim (5,-2)$ and $\chi \sim (7,0)$ are totally symmetric tensors 
$\Phi_{abcd}$  and $\chi_{abcdef}$ providing a number of multiply-charged component states 
\begin{eqnarray}
    \begin{matrix} \Phi_{1111} = \phi^+ \\ \Phi_{1112} = \frac{1}{\sqrt{4}}\phi^{0} \\ \Phi_{1122} =  \frac{1}{\sqrt{6}}\phi^{-} \\ 
    \Phi_{1222} = \frac{1}{\sqrt{4}}\phi^{--}  \\ \Phi_{2222} =  \phi^{---}   \end{matrix}   \ \ ,
&& \begin{matrix} \chi_{111111} = \chi^{+++} \\ \chi_{211111} = 
    \frac{1}{\sqrt{6}}\chi^{++} \\ \chi_{221111} = \frac{1}{\sqrt{15}}\chi^{+}\\ 
    \chi_{222111} = \frac{1}{2\sqrt{5}}\chi^{0}  \\ \chi_{222211} = 
    \frac{1}{\sqrt{15}}\chi^- \\ \chi_{222221} = 
\frac{1}{\sqrt{6}}\chi^{--} \\ \chi_{222222} = \chi^{---} \end{matrix} \ \ ,
\label{components}
\end{eqnarray}
where we distinguish $\phi^{-}$ and $(\phi^{+})^*$ for the complex scalar.

The full scalar potential contains gauge invariant pieces 
\begin{eqnarray}\label{scalarpot}
\nonumber  V(H_{\bf{1}},H_{\bf{2}},\Phi, \chi) &=& V(H_{\bf{1}},H_{\bf{2}}) + V(\Phi)  + V(\chi) + V_m(\Phi,\chi) \\ 
\nonumber   &+&  V_m(H_{\bf{1}},H_{\bf{2}},\Phi) + V_m(H_{\bf{1}},H_{\bf{2}},\chi)    \\
            &+&  V_m(H_{\bf{1}},H_{\bf{2}},\Phi, \chi)  \ ,
\end{eqnarray}
where the first term $V(H_{\bf{1}},H_{\bf{2}})$ is explicated in (\ref{2HDpot}) and we will not need explicit form of the terms $V(\Phi)$, $V(\chi)$ and $V_m(\Phi,\chi)$ in this paper. The terms $V_m(H_{\bf{1}},H_{\bf{2}},\Phi)$ and  $V_m(H_{\bf{1}},H_{\bf{2}},\chi)$ are important for the diphoton signal and therefore will be introduced later in (\ref{scalarpot-chi}) and (\ref{scalarpot-Phi}). Finally, the last term is relevant for neutrino mass and will be discussed next.\\

\underline{\bf Neutrino mass:}
This last term  represents  the $\tilde Z_2$-symmetric mixing potential
\begin{eqnarray}\label{NEWdim4}
  V_m(H_{\bf{1}},H_{\bf{2}},\Phi, \chi) &=&  \kappa H_{\bf{1}} H_{\bf{2}} \Phi \chi + \mathrm{h.c.} \ ,
\end{eqnarray}
which provides the couplings needed to close the three-loop neutrino mass diagram. After EWSB, the relevant 2HDM piece undergoes the substitution:
\begin{eqnarray}\label{HHPhichi-vertex}
\kappa (H^{+}_{\bf{1}} H^{0}_{\bf{2}} +  H^{+}_{\bf{2}} H^{0}_{\bf{1}}) \,  \rightarrow  \, v\, \kappa  \, \cos 2\beta\,   H^{+}\ .
\end{eqnarray} 
so that the resulting quartic vertices together with the appropriate Yukawa couplings
\begin{eqnarray}\label{yukawa}
 {\cal L}_Y
   = - y_{e_i}  \overline{L}_{i L} H_{\bf{1}} e_{i R}
     - g_{i \alpha}  \overline{(e_{i R})^c} {\Phi}^* \Sigma_{\alpha R} + \mathrm{h.c.} \ .
\end{eqnarray}
complete the neutrino mass diagram. In our lepton-specific 2HDM, only the Higgs doublet $H_{\bf{1}}$ couples to the SM leptons, so that 
the SM lepton mass $m_e$ corresponds to the Yukawa strength $y_{e_i}^{SM}
=y_{e_i} v_1/v=\sqrt{2}m_{e_i}/v$.

Collecting all the pieces, we finally arrive at the resulting three-loop-generated lepton-number-breaking Majorana neutrino
mass matrix $M^\nu_{ij}$ for active neutrinos, 
which keeps the form of~\cite{Aoki:2009vf} and reads
\begin{eqnarray}
M_{ij}^\nu &=& \sum_{\alpha=1}^3 
   C_{ij}^\alpha \, F(m_{H^\pm}^{},m_\Phi^{}, m_\chi, m_{\Sigma_{\alpha}}) \label{eq:mij} \ .
   \label{neutrino3loop}
\end{eqnarray}
Here the coefficient $C_{ij}^\alpha$ comprises the vertex coupling strengths
\begin{eqnarray}
C_{ij}^\alpha &=&
    \frac{7}{3} \kappa^2  \tan^2\beta \cos^2 2\beta \,
  y_{e_i}^{\rm SM} g_i^\alpha y_{e_j}^{\rm SM} g_j^\alpha, 
\end{eqnarray}
and the loop integral is represented by the function $F$,
expressed in terms of the Passarino-Veltman function
for one-loop integrals~\cite{passarino-veltman}.
In the wide range of the parameter space, the magnitude of $F$ is of the order $10^{2}$ eV so that (\ref{neutrino3loop}) reproduces the neutrino masses  with the coefficient  $C_{ij}^\alpha \leq 10^{-4}$ that is easily 
achieved with natural values of $\mathcal{O}(1)$ for the couplings of the model.

\section{Constraints from the diphoton signals} \label{diphoton}

\subsection{The one-loop model}

After EWSB, the neutral components of the SM Higgs doublet $\phi^0$ and the triplet $\Delta^0$ mix with an angle $\theta_0$, yielding  
$h(125)$ and $H(750)$ candidates. 
As discussed above, the VEV for the neutral triplet component is constrained by electroweak measurements to be $\langle \Delta^0\rangle <\mathcal{O}$(1) GeV so that we neglect effects of $\mathcal{O}(\langle \Delta^0\rangle/v)$. We also take the  quartic coupling $\lambda_7\simeq 10^{-4}$  as deduced from the neutrino masses in Sec.~2.1.
There are also charged components of  the triplet $\Delta^\pm$ and the charged scalar $h^+$ which enter into quantum loops relevant for production and decays of the light  SM-like Higgs $h(125)$ and its heavy relative $H(750)$.\\
 
\begin{figure}
\centerline{\includegraphics[scale=0.68]{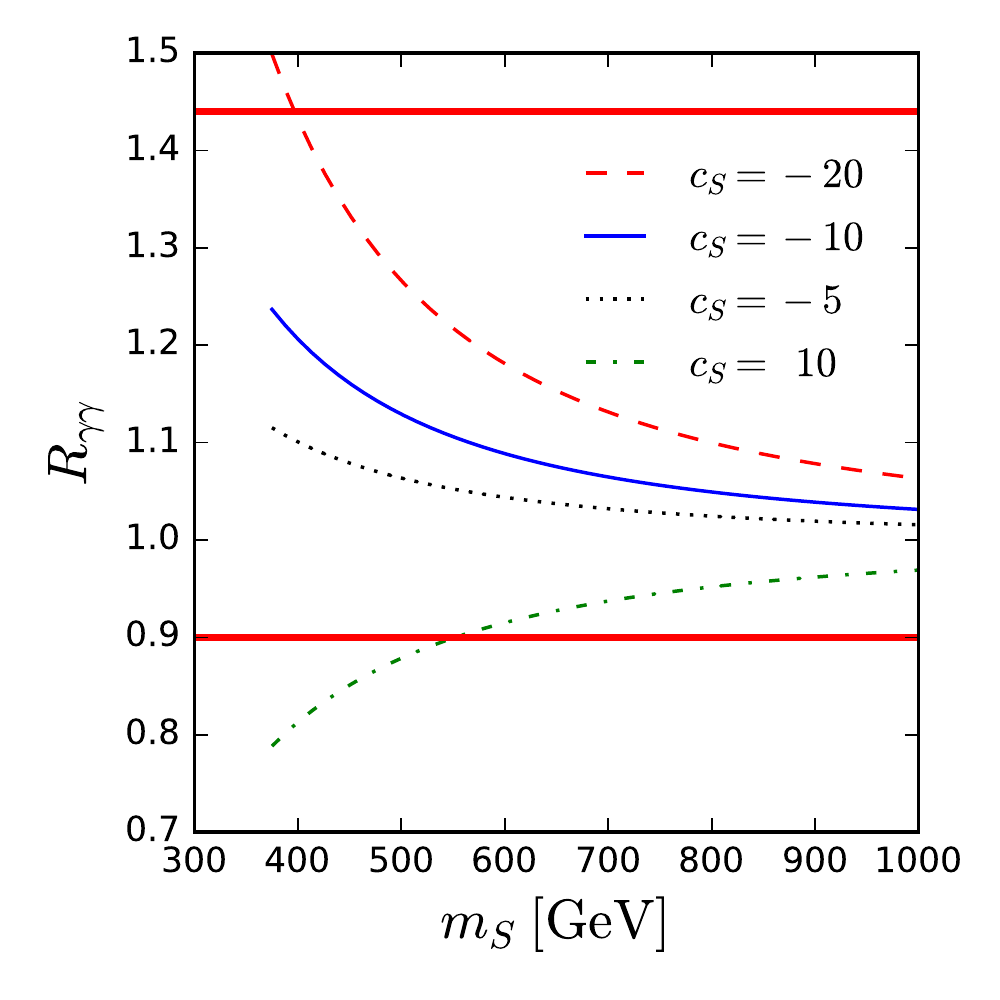}\includegraphics[scale=0.68]{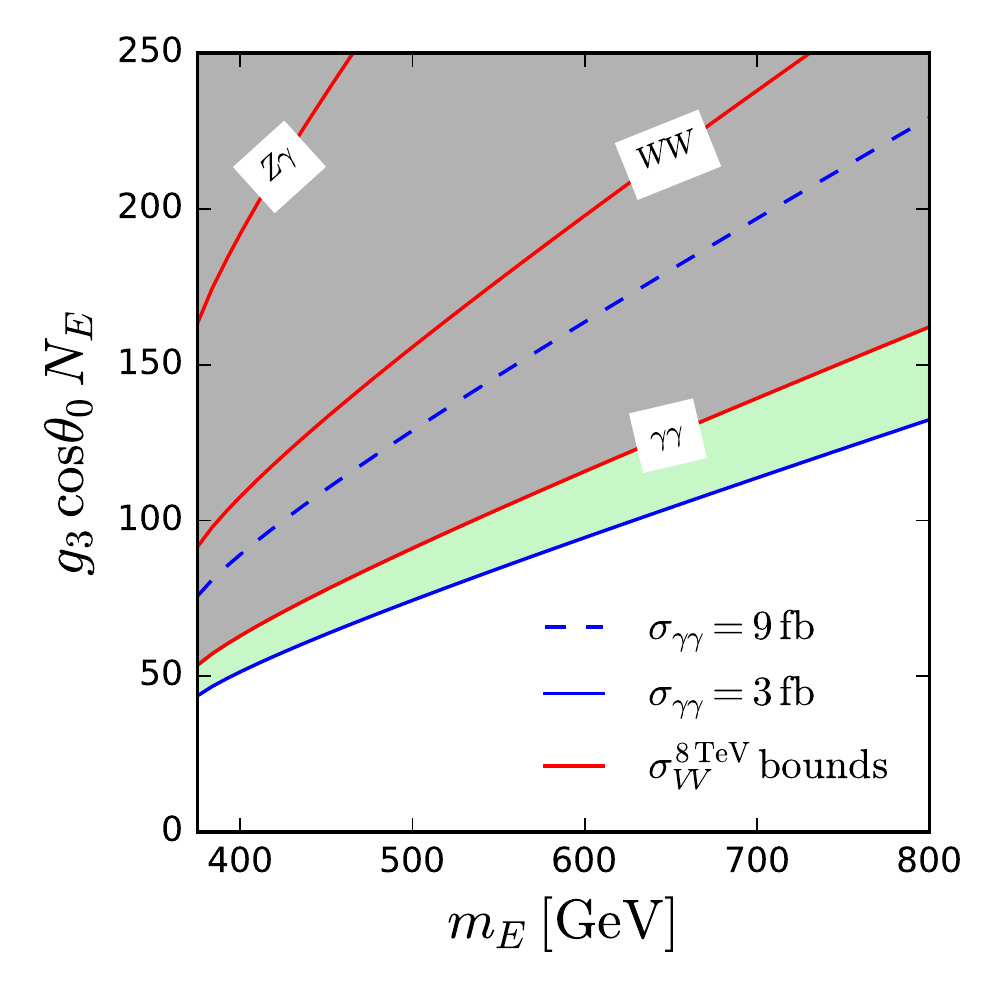}}
\caption{\small Enhancement factor $R_{\gamma\gamma}$ for the $h \rightarrow \gamma \gamma$ decay width in dependence on scalar coupling $c_S$ and the
mass $m_S$ of the lighter charged scalar (left). Region of parameter space where
one-loop model explains 750 GeV diphoton resonance (light/green) allowed (dark/grey) by the
LHC 8 TeV constraints (right).
}
\label{rgamma}
\end{figure}

\underline{\bf The 125 GeV Higgs:} 
For a sole hypercharge-zero scalar triplet extension of the SM,
studied previously in detail in~\cite{FileviezPerez:2008bj}, the LHC diphoton signal has been studied in~\cite{Wang:2013jba}.
For the one-loop model at hand, containing additional charged singlet scalar $h^+$,
 we extend for completeness the previous study of the diphoton signal~\cite{Brdar:2013iea} to new mass region of charged BSM scalars in the loop, as motivated by the recently hinted  750 GeV resonance.
As in~\cite{Brdar:2013iea},
the scalar $h(125)\simeq \phi^0\cos\theta_0$ is predominantly given by the neutral component of the SM Higgs doublet $\phi^0$, which couples via $c_S v \phi^0 S^\dagger S$  to BSM charged scalars $S(h^+,\Delta^+)$, and they  in loop contribute to diphoton decay amplitude. Thereby, the $c_S$ couplings are linked to the couplings  $\lambda_4$ and $\lambda_5$ in (\ref{potential}). In the conventions and notations 
from~\cite{Carena:2012xa,Picek:2012ei}, the enhancement factor with respect to the SM decay width
is displayed in the left panel of Fig.~\ref{rgamma}. The horizontal lines in this figure highlight the current bound $R_{\gamma\gamma} = 1.17\pm 0.27$~\cite{Aad:2015gba}.
Since the contribution of the lighter among the two charged scalars $S$ dominates, this figure sets a bound on the respective coupling $c_S$.\\ 
Only the charged scalars which are sufficiently light may produce significant effects in the LHC diphoton Higgs signals, so that there is poor constraint  on $c_S$ couplings of the charged scalars which exceed a half of mass of the $H(750)$ scalar particle.\\

\underline{\bf The 750 GeV scalar:} Here we attempt to fit the heavy state $H\simeq \Delta^0\cos\theta_0 $, which is predominantly $\Delta^0$ in this model, to the hinted  $H(750)$ scalar particle. Let us first discuss the productions mechanisms for $H(750)$. For $\langle \Delta^0\rangle=0$ there is no tree-level coupling of $H$ to the SM fermions and vector bosons\footnote{In general $g_{Hff}$ and $g_{HVV}$ are $\sim \sin\theta_0\sim \frac{\langle \Delta^0\rangle}{v}$ which is small. However, if $2M_{\Delta^+}^2=M^2_{H(750)}+M^2_{h(125)}$ the mixing can become sizeable \cite{FileviezPerez:2008bj}. We assume this does not happen here.} and therefore the gluon fusion production is negligible. 
We are thus led to consider the EW vector boson fusion (VBF) mechanisms. 
For resonance much heavier than electroweak scale, photon fusion production
mechanism dominates and we neglect the contributions from fusion of
weak bosons (see discussion in \cite{Fichet:2015vvy}).
The diphoton signal strength at
$\sqrt{s}=13$ TeV from the photon fusion is given by  \cite{Fabbrichesi:2016alj}:
\begin{eqnarray}
\sigma_{\gamma\gamma}\equiv\sigma (pp \to H\to \gamma \gamma)=10.8\, {\rm pb}\times \frac{\Gamma_H}{45 \, {\rm GeV}}\times {\rm Br}(H\to \gamma\gamma)^2 \;,
\label{gammafusion}
\end{eqnarray}
where we account for the photoproduction that includes both elastic and
inelastic contributions \cite{Csaki:2016raa}. 
To estimate the contributions of charged scalars to the one-loop generated $H\gamma \gamma$ coupling, we use Lagrangian (\ref{potential}). Here we notice that the leading trilinear couplings $\lambda_{H \Delta^+ \Delta^-}\sim  \lambda_3 \langle \Delta^0\rangle   \cos \theta_0$ and $\lambda_{H h^+ h^-}\sim  \lambda_6 \langle \Delta^0\rangle  \cos \theta_0$ relevant for the charged scalar loop
vanish in the limit $\langle \Delta_0\rangle=0$ and the remaining quartic couplings are negligible. We therefore need to consider the charged fermion loops and the leading contribution from Yukawa couplings in (\ref{Yukawa1loop}) is  represented by $g_3 \bar{E}_L \Delta E_R+h.c.$ term. 
The vector-like fermion loop-generated couplings of 750 GeV candidate to different channels with SM gauge bosons for the degenerate coupling $\lambda$ read~(\emph{e.g.} \cite{Cao:2015apa}):
 
\begin{align}
g_{H\gamma\gamma} &= \lambda \alpha\sum_F  \left\lbrace  Q_F^2 \right\rbrace  \frac{S_{1/2}(\tau_F)}{m_F} \;, \nonumber \\
g_{HZ\gamma} &=  \lambda\alpha\sum_F \left\lbrace \sqrt{2} Q_F\frac{(T_{3F}-s_W^2
Q_F)}{s_W c_W} \right\rbrace \  \frac{S_{1/2}(\tau_F)}{m_F}\;, \nonumber \\ 
g_{HZZ} &= \lambda\alpha
\sum_F  \left\lbrace \frac{(T_{3F}-s_W^2 Q_F)^2}{s_W^2 c_W^2}  \right\rbrace  \frac{S_{1/2}(\tau_F)}{m_F} \;, \nonumber \\
g_{HWW} &=  \lambda\alpha\sum_F  \left\lbrace \sqrt{2}
\frac{(T_F-T_{3F})(T_F+T_{3F}+1)}{2 s_W^2} \right\rbrace 
\frac{S_{1/2}(\tau_F)}{m_F} \;.
\label{effcoupl}
\end{align}
Here, $T_F$ is the weak isospin of the loop-fermion $F$, the triangle loop function is given by $S_{1/2}(\tau_F)= 2\tau_F (1+ (1-\tau_F )\arcsin^2(1/\sqrt{\tau_F}))$, and the respective variable is $\tau_F=4 m_F^2/M_H^2$. The couplings include
symmetrization factors for identical particles in the final state, and are
normalized so that, neglecting masses of the $W$ and $Z$ bosons give:
\begin{eqnarray}
\label{finalphotonF}
\Gamma(H \rightarrow VV)=\frac{M_H}{64\pi^3}\bigg|\frac{M_H\, g_{HVV}}{2}\bigg|^2 \ . 
\end{eqnarray} 
For degenerate loop masses, the couplings can be compactly expressed in terms
of quadratic Dynkin indices $I_1$ and $I_2$ of the loop-fermion SM group representations:
\begin{gather}
g_{H\gamma\gamma} = \lambda \alpha (I_1+I_2) \frac{S_{1/2}(\tau_F)}{m_F} \,,  \qquad 
g_{HZ\gamma} = \sqrt{2} \lambda \left(\frac{c_W}{s_W}I_2 - \frac{s_W}{c_W}I_1\right)
\frac{S_{1/2}(\tau_F)}{m_F}\,, \nonumber \\ 
g_{HZZ} = \lambda\alpha\left(\frac{c_{W}^2}{s_{W}^2}I_2 + \frac{s_{W}^2}{c_{W}^2}I_1\right)
\frac{S_{1/2}(\tau_F)}{m_F}  \,, \quad
g_{HWW} = \sqrt{2} \lambda\alpha \frac{I_2}{s_W^2} \frac{S_{1/2}(\tau_F)}{m_F} \,.
\label{effcouplD}
\end{gather}
For the vector-like fermion $E_{L,R}$ at hand with multiplicity $N_E=3$,
we have $\lambda=g_3 \cos\theta_0 N_E$, $I_1=1/2$, $I_2=1/2$. For the resulting ratio of the decay widths
\begin{equation}
    R_{VV} \equiv \frac{\Gamma(H\to VV)}{\Gamma(H\to \gamma\gamma)}
    \label{eq:RVV},
\end{equation}
we obtain
\begin{eqnarray}
R_{WW} \approx 9.1 \,, \quad
R_{ZZ} \approx 3.2 \,,  \quad
R_{Z\gamma} \approx 0.8 \,.
\end{eqnarray}
This results in a branching ratio $Br(H\to \gamma\gamma) \approx 7\,\% $. 
Comparing to the diphoton signal strength (\ref{gammafusion}), one can explain the diphoton resonance
with cross-section of 3-9 fb by using narrow width of the resonance $\Gamma_H \sim 2.5-7.5\,{\rm GeV}$.

In this narrow width scenario, leading to $\Gamma(H\to\gamma\gamma) = 0.18-0.53\,{\rm GeV}$,
we can now investigate the influence of the constraints coming
from the searches for resonances decaying to gauge boson pairs at the LHC 8 TeV run. Constraints on the cross sections 
$\sigma_{VV}^{8\,{\rm TeV}}\equiv \sigma(pp \to H \to VV)$ are 
\cite{CMS:2014onr,Aad:2015mna,Aad:2014fha,Aad:2015kna,Khachatryan:2015cwa,Aad:2015agg}
\begin{equation}
    \sigma_{\rm WW}^{8 \rm  TeV}< 40 \,{\rm fb}\,, \quad 
    \sigma_{\rm ZZ}^{8 \rm  TeV}< 12 \,{\rm fb}\,, \quad 
    \sigma_{\rm Z\gamma}^{8 \rm  TeV}< 11 \,{\rm fb}\,, \quad 
    \sigma_{\rm \gamma\gamma}^{8 \rm  TeV}< 1.5 \,{\rm fb} \,.
\label{run1}
\end{equation}
To make a comparison between 8 TeV data (always explicitly indicated) and 13 TeV data,
we need the value for the gain ratio $r_{\gamma\gamma}$
of the photon fusion production cross-sections at 13 TeV and at 8 TeV,
so that
\begin{equation}
    \sigma_{VV}^{8\,{\rm TeV}} = \frac{\sigma_{\gamma\gamma}}{r_{\gamma\gamma}} R_{VV} \,.
    \label{eq:VV8}
\end{equation}
This gain ratio is often taken to be 
$r_{\gamma\gamma}\approx 2$ \cite{Franceschini:2015kwy}, which would create 
strong tension with non-observation of the diphoton resonance in the 8 TeV LHC data. 
However, more elaborate analyses \cite{Fichet:2015vvy,Csaki:2015vek}, 
taking into account also elastic emission of the photon as well as finite proton
size effects lead to increased ratios up to 3.9, alleviating this tension. We take average value of $r\approx 3$ 
and obtain for the  $\sigma_{\gamma\gamma} = 3-9\,{\rm fb}$ range
\begin{equation}
    \sigma_{\rm WW}^{8 \rm  TeV} = 9-27\,{\rm fb}\,, \;
    \sigma_{\rm ZZ}^{8 \rm  TeV} = 3-10 \,{\rm fb}\,, \;
    \sigma_{\rm Z\gamma}^{8 \rm  TeV} = 0.8-2.4 \,{\rm fb}\,, \;
    \sigma_{\rm \gamma\gamma}^{8 \rm  TeV} = 1-3 \,{\rm fb}\,.
\label{eq:pred8}
\end{equation}
We see that the LHC 8 TeV run constraint on $\sigma_{\rm \gamma\gamma}^{8 \rm  TeV}$ is violated for parameters
corresponding to larger values
of $\sigma_{\gamma\gamma}$, and that $\sigma_{\gamma\gamma} \sim 3-4.5\,{\rm fb}$
is preferred. Even in this case,
one expects that additional gauge boson pairs from hinted 750 GeV resonance
should show up soon as more data are gathered in the LHC 13 TeV run.
Results above are summarized in Fig.\ref{rgamma}(right),
showing that, most importantly, for the dominant portion of the parameter space this model requires either non-perturbative value of the
coupling $g_3 > 4\pi$ or larger multiplicities $N_E>3$. For $N_E=3$, the value $g_3\approx 4\pi$ is achieved only for $m_E\approx 375$ GeV.

We might improve the capacity of our model to account for a diphoton excess by introducing appropriate coloured 
degrees of freedom~\cite{Bertuzzo:2016fmv}.
Numerous models employed a vector-like singlet quark to enhance the production cross section.
In the present case it amounts to  extending the radiative model~\cite{Brdar:2013iea} to the quark-lepton symmetric
version containing the vector-like top-partner.
Comparing to relatively weak bounds for charged and neutral leptons,
typically around 100 GeV \cite{Agashe:2014kda}, 
the corresponding limits for new heavy charge-2/3 quarks are 720-920
GeV~\cite{Khachatryan:2015oba}
and 715-950 GeV~\cite{Aad:2015kqa}.\\
Instead of trying to reproduce the diphoton excess with beyond SM fermions we can try to employ
higher electroweak scalar multiplets containing a plethora of charged states. Such scenario is offered in a recent 
three-loop neutrino mass model~\cite{Culjak:2015qja}, which we consider in the next section.

\subsection{The three-loop model}
In this model the hinted $H(750)$ scalar particle is the heavy CP-even neutral scalar emerging from the 2HDM. The $H(750)$ state does not couple to exotic quintuplet fermion $\Sigma$ in gauge invariant way. We therefore consider the contributions to diphoton signal from the exotic charged scalar particles contained in fields $\Phi$ and $\chi$ defined in (\ref{components}). 

Let us start with quartic vertices which generate triangle loops with exotic charged scalars for diphoton decays. 
These couplings  can be read from the scalar potentials contained in (\ref{scalarpot}):
\begin{eqnarray}\label{scalarpot-chi}
  V_m(H_{\bf{1}},H_{\bf{2}},\chi) &\supset& (\tau_1H_{\bf{1}}^\dagger H_{\bf{1}} +  \tau_2 H_{\bf{2}}^\dagger H_{\bf{2}}) \chi^\dagger \chi \ ,
\end{eqnarray}
and
\begin{eqnarray}\label{scalarpot-Phi} V_m(H_{\bf{1}},H_{\bf{2}},\Phi) \supset  (\sigma_1H_{\bf{1}}^\dagger H_{\bf{1}}+\sigma_2H_{\bf{2}}^\dagger H_{\bf{2}}) \Phi^\dagger \Phi +  (\sigma'_1H_{\bf{1}}^* H_{\bf{1}}+ \sigma'_2H_{\bf{2}}^* H_{\bf{2}}) \Phi^* \Phi . \end{eqnarray}
We start with (\ref{scalarpot-chi}) where the trilinear couplings  strengths $h(125)\chi^\dagger \chi$ and $H(750)\chi^\dagger \chi$
are extracted after using the VEVs $v_1=v\  \mathrm{cos}\beta$ and $ v_2=v \ \mathrm{sin}\beta$ in one of the doublets. 
This substitution leads to the universal coupling for all charged components of real scalar septuplet $\chi$ to $h(125)$ and $H(750)$
\begin{eqnarray}\label{HHchi-tau1,2-vertex}
&V_{\chi}&= (\tau_1 H^{0}_{\bf{1}} H^{0}_{\bf{1}} + \tau_2 H^{0}_{\bf{2}} H^{0}_{\bf{2}})\chi^\dagger \chi  =  \\
     \nonumber        &=&v\chi^\dagger \chi \bigg[H (\tau_1 \mathrm{cos}\alpha \ \mathrm{cos}\beta
+ \tau_2  \mathrm{sin}\alpha\ \mathrm{sin}\beta )+h (- \tau_1 \mathrm{sin}\alpha\ \mathrm{cos}\beta
+ \tau_2  \mathrm{cos}\alpha \ \mathrm{sin}\beta ) \bigg]\ .
\end{eqnarray}
By working in the following in the ``alignment limit" of the 2HDM~\cite{Han:2015qqj}
\begin{eqnarray}\label{alignment-constr-a-b}
\mathrm{tan}\beta = 1  \   \  \ , \ \   \mathrm{sin}(\beta-\alpha) = 1  \ ,
\end{eqnarray}
and assuming that couplings satisfy the relation 
\begin{eqnarray}\label{alignment-constr-tau}
\tau_1 = -\tau_2\equiv \tau 
\end{eqnarray}
will lead us to
\begin{eqnarray}\label{HHchi-tilde-tau-vertex}
 V_{\chi}=  \ v \  
\tau \ \big[\mathrm{cos}(\beta+\alpha)   \   H -\mathrm{sin}(\beta+\alpha) \   h\big]\ \chi^\dagger \chi=v \ \tau \  H\chi^\dagger \chi.
\end{eqnarray}
This alignment limit identifies the light state as SM-like $h(125)$, such that
its diphoton decay acquires no contribution  from (\ref{HHchi-tilde-tau-vertex}). Explicitly, the couplings of the charged components of the septuplet to $H(750)$ are :
\begin{eqnarray}
    V_\chi =  \tau \ v  \ (\chi^{+} \chi^{-} +  \chi^{++} \chi^{--} 
                  +  \chi^{+++} \chi^{---})  \   H  \ .
\end{eqnarray}

The septuplet scalar components are degenerate at the tree-level
\begin{eqnarray}\label{spectrum-chi}
  m^2_\chi &=& \mu_\chi^2 + \frac{\tau}{2} v^2(\mathrm{cos}^2\beta - \mathrm{sin}^2\beta)\ ,
\end{eqnarray}
where the EWSB correction vanishes for $\tan\beta=1$.

Similarly, for the quintuplet $\Phi$, we impose equivalent conditions on the couplings in (\ref{scalarpot-Phi}),
\begin{eqnarray}\label{alignment-constr}
\sigma_1 = - \sigma_2\equiv \sigma  \   \  \ , \ \  \sigma'_1 = - \sigma'_2\equiv \sigma'   \ ,
\end{eqnarray}
so that the trilinear couplings of $h(125)$ to the charged components of the quintuplet vanish. The $H(750)$ couplings to these charged components of the quintuplet, relevant for the $H \rightarrow \gamma \gamma$ decay, are
\begin{eqnarray}   V_\Phi =v H  (c_{\Phi^{+}} \Phi^{+*} \Phi^{+} + c_{\Phi^{-}} \Phi^{-*} \Phi^{-}
                  +  c_{\Phi^{--}} \Phi^{--*} \Phi^{--} + c_{\Phi^{---}} \Phi^{---*} \Phi^{---})  \,
\end{eqnarray}
where the newly introduced couplings simplify according to (\ref{alignment-constr}) as
\begin{eqnarray}\label{couplings-Phi-H}
  c_{\Phi^{+}}= \sigma, \;
  c_{\Phi^{0}}= \sigma + \frac{\sigma'}{4} ,\; 
  c_{\Phi^{-}}= \sigma + \frac{\sigma'}{2} ,\;
  c_{\Phi^{--}}= \sigma + \frac{3\sigma'}{4} ,\; 
  c_{\Phi^{---}}= \sigma +  \sigma' . 
\end{eqnarray}
In contrast to septuplet case, the EWSB contributions to the mass of different components of the complex quintuplet $\Phi$ 
are not the same and are given as
\begin{eqnarray}\label{spectrum-Phi}
  m^2_{\Phi^{(Q)}} = \mu_\Phi^2 + \frac{1}{2} v^2(\mathrm{cos}^2\beta - \mathrm{sin}^2\beta) 
 c_{\Phi^{(Q)}} \ . 
\end{eqnarray}

\begin{figure}[th]
\centerline{\includegraphics[scale=0.64]{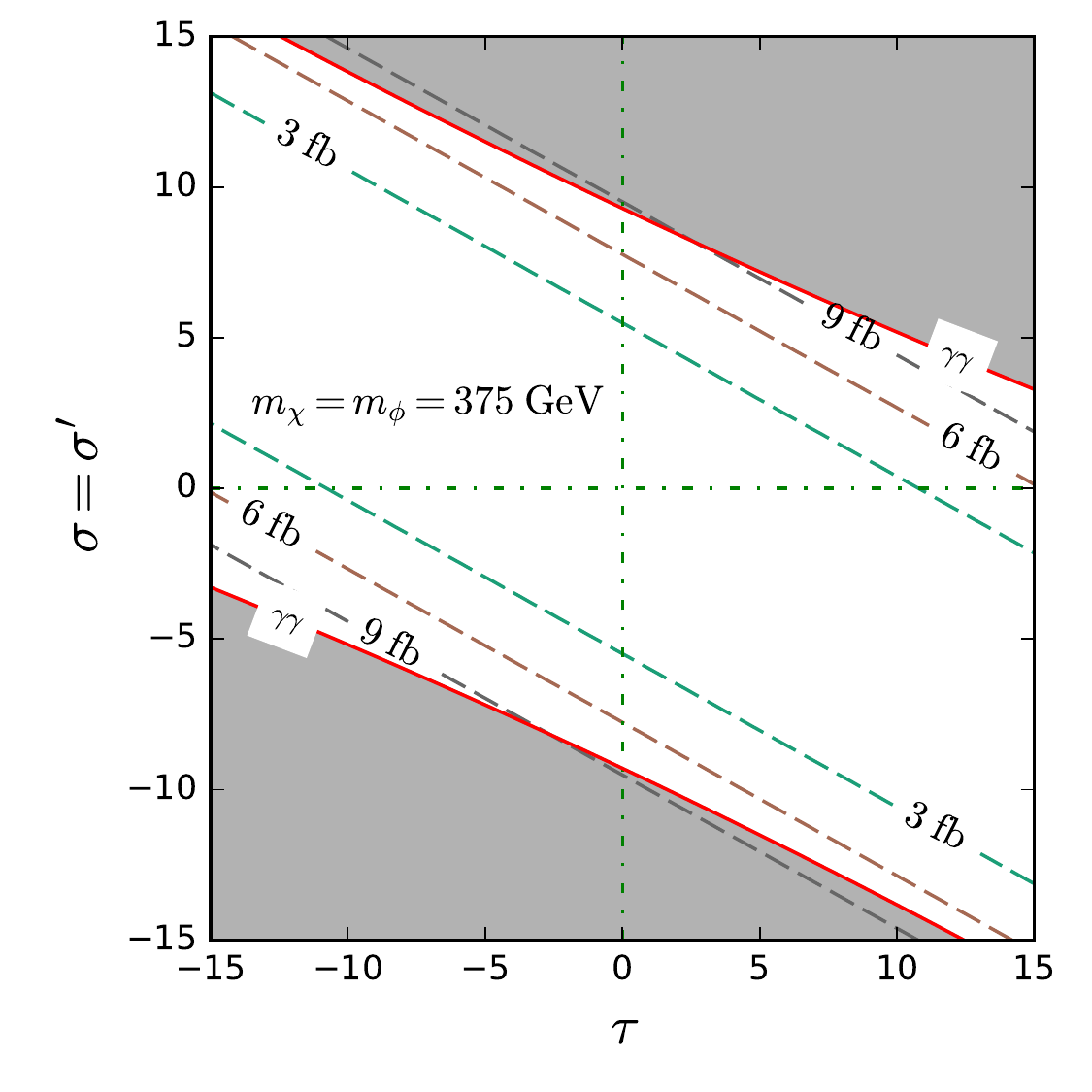}%
\includegraphics[scale=0.64]{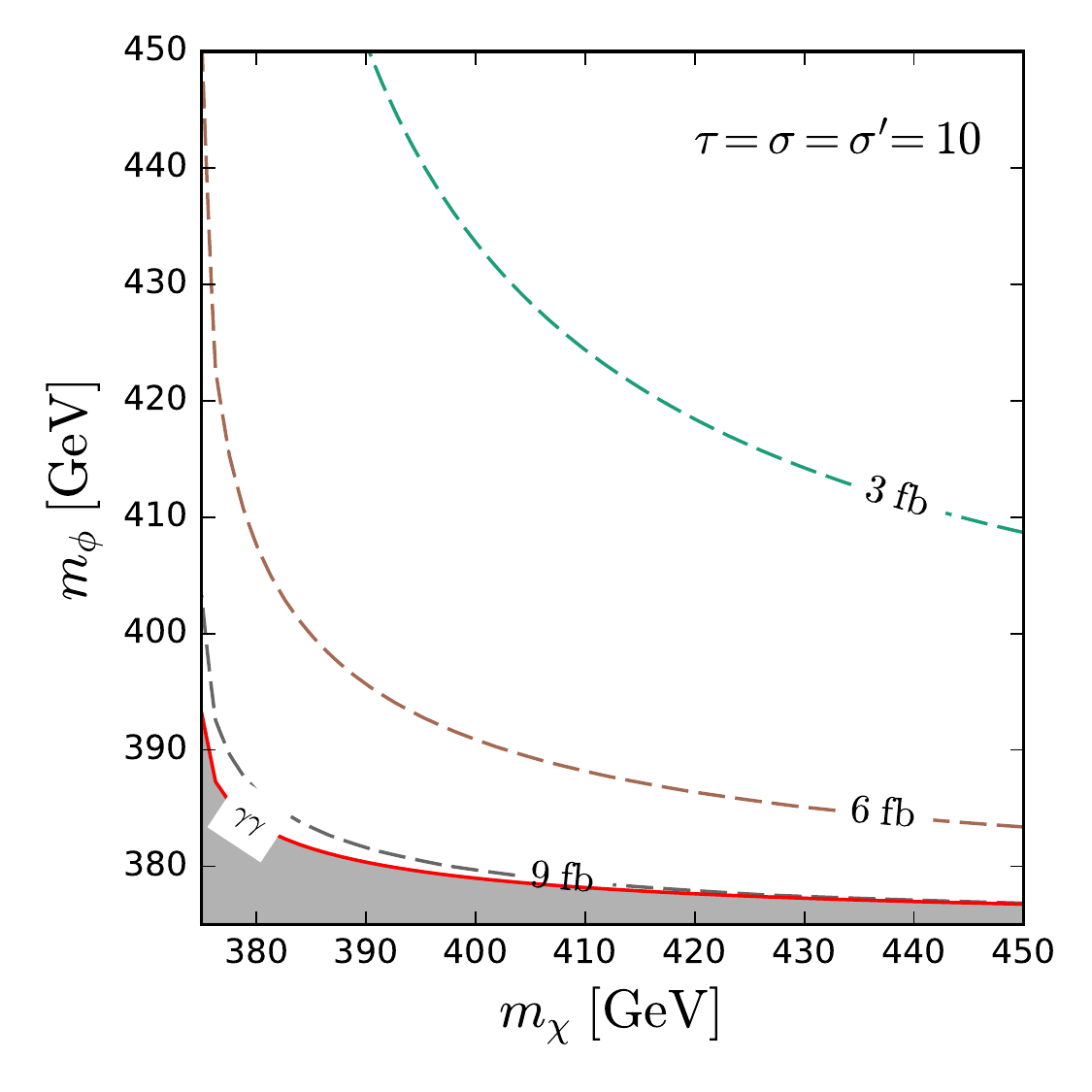}}
\caption{\label{fig:param-ranges} 
Cross section for $p p \to H(750) \to \gamma\gamma$ (dashed lines)
in the three-loop neutrino mass model for ranges of values for
coupling (left) and mass parameters (right). Grey area is excluded by LHC search for  $p p \to H \to \gamma\gamma$
at 8 TeV.
}
\end{figure}
Again for $\tan\beta=1$ the EWSB contributions vanish.\\
In the three-loop model, the diphoton excess may be explained by the gluon-fusion production process of $H$ and $A$. In the lepton-specific 2HDM at hand, only $H_2$ couples to the SM quarks and the relevant couplings of $H(750)$ in the alignment limit are given by ($V=W^\pm, Z)$ \cite{Bertuzzo:2016fmv}:
\begin{eqnarray}
\frac{g_{Htt}}{g_{Htt}^{SM}} &=& \cos(\beta-\alpha)-\frac{\sin(\beta-\alpha)}{\tan\beta}=-1 \qquad \quad
\frac{g_{Att}}{g_{Htt}^{SM}} = \frac{1}{\tan\beta} =1 \nonumber \\ 
g_{HVV}&=& 2 \cos(\beta-\alpha)\frac{m_V^2}{v}=0 \ .
\label{alignment}
\end{eqnarray}\\
The loop of the quintuplet and septuplet charged scalar states contributes only to the decay of the CP-even $H$ boson, so that the decay rate of
$A$ into diphoton is not enhanced by these charged scalar loops. The dominant coupling of $A$ is to the top quark and for $A t\bar{t}$ taking the SM $Ht\bar{t}$ value, this coupling mediates the $\sigma (pp \to A\to \gamma \gamma)\sim 0.01$ fb which is about $1000$ times smaller than required to explain the diphoton excess. For this reason we do not consider the contribution of the $A$ state to the diphoton signal further. For the same reason we neglect the decay of $H$ through the top-quark loop in the estimate of $\sigma (pp \to H\to \gamma \gamma)$ and keep only the decay through the new charged states.
 
 In the scenario where $H$ is produced dominantly through
    gluon-gluon fusion, diphoton cross section is
    \begin{equation}
    \sigma_{\gamma\gamma}  = \sigma_{ggF} Br(H\to\gamma\gamma)\;,
    \end{equation}
    where cross section for $pp\to gg X \to H X$
    is $\sigma_{ggF} = 737\,{\rm fb}$
    at $\sqrt{s} = 13\,{\rm TeV}$, and $\sigma_{ggF}^{8\,{\rm TeV}} = 
    157\,{\rm fb}$
    at $\sqrt{s} = 8\,{\rm TeV}$ \cite{Heinemeyer:2013tqa}.

\begin{figure}[th]
\centerline{\includegraphics[scale=0.68]{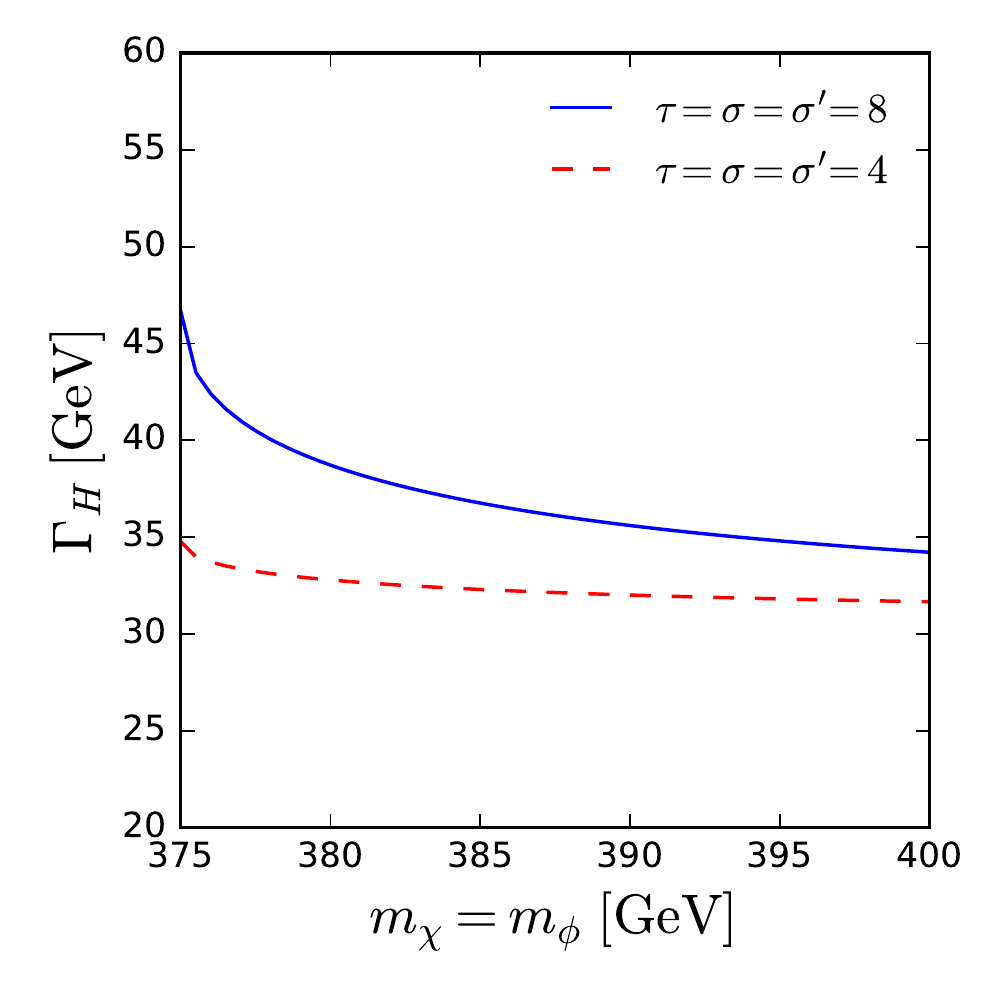}%
\includegraphics[scale=0.68]{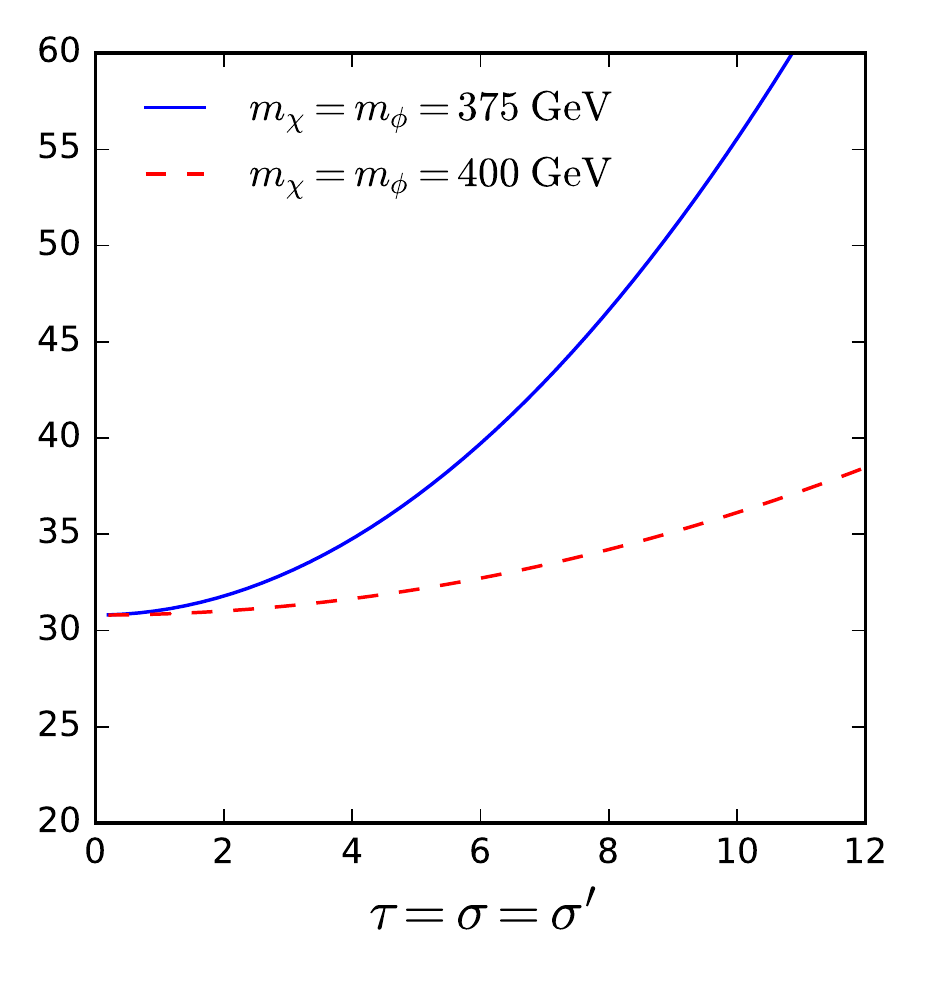}}
\caption{\label{fig:GAMTOT} The total decay width of $H(750)$ particle
in the three-loop neutrino mass model for the generic choice of the parameters.
}
\end{figure}

The decay width of the $H(750)$ to the $\bar{t}t$ pair is:
\begin{equation}
\Gamma(H\to t\bar{t})= N_c  \frac{\alpha M_H}{8 \sin^2\theta_W} \frac{m_t^2}{m_W^2} \bigg(1-\frac{4 m_t^2}{m_H^2}\bigg)^{3/2} \approx 30 \ \rm {GeV}
\label{eq:Htt}
\end{equation}
which is roughly what is observed by ATLAS \cite{Aaboud:2016tru}. We therefore take the masses of the new charged scalar states to be $\mu_{\chi,\Phi}>$ 375 GeV as otherwise the decay channel of $H(750)$ to these states opens up and the resonance quickly becomes very wide.

Additional subleading contributions to the $H(750)$ width are provided by the decays into SM vectors. In the alignment limit, the tree-level couplings $H(750) VV$, from (\ref{alignment}), are absent so that these decay modes are generated only at one-loop level. 
Again, it is convenient to introduce the effective couplings $g_{HVV}$ of $H(750)$ to the
SM gauge bosons. They can be obtained from those in (\ref{effcoupl}) with substitutions of the corresponding terms in curly braces
\begin{equation}
\lambda\sum_F \left\lbrace  \cdots\right\rbrace \frac{A_{1/2}(\tau_F)}{m_F}  \longrightarrow  
\tau   \sum_S \left\lbrace \cdots\right\rbrace \frac{v A_{0}(\tau_S)}{2 m_{S}^2} \;,
\label{effcouplS7}
\end{equation}
for the real septuplet contribution, and
\begin{equation}
\lambda\sum_F\left\lbrace  \cdots\right\rbrace \frac{A_{1/2}(\tau_F)}{m_F}  \longrightarrow  
\sum_S \left(\sigma + \sigma'\Big(\frac{2-T_{3S}}{4}\Big)\right) \left\lbrace \cdots \right\rbrace
\frac{v A_{0}(\tau_S)}{m_{S}^2} \;,
\label{effcouplS5}
\end{equation}
for the complex quintuplet. Here the factor  $(2-T_{3S})/4$ accounts for
the non-universality of coupling to $H$ (\ref{couplings-Phi-H}),
and should be changed to $(3-T_{3S})/8$ in the sole case of $g_{HWW}$.
These constants are normalized so that, neglecting masses of the $W$ and $Z$ bosons,
\begin{eqnarray}
\label{finalphotonS}
\Gamma(H\to VV)=\frac{M_H}{256\pi^3}\bigg|\frac{M_H\, g_{HVV}}{2}\bigg|^2 \ . %\frac{g_{H\gamma\gamma}^2}{64\pi} m_H^3
\end{eqnarray}
The variable $\tau_S\equiv 4 m_S^2/m_H^2$ and the loop function is given by
$A_0(\tau_S)\equiv-\tau_S (1-\tau_S \arcsin^2 (1/\sqrt{\tau_S}))$.
For the degenerate couplings $\tau=\sigma=\sigma'$, 
this leads to the ratios of diboson to diphoton decay widths (\ref{eq:RVV})
\begin{eqnarray}
R_{WW} \approx 17.8 \,, \quad
R_{ZZ} \approx 4.9 \,,  \quad
R_{Z\gamma} \approx 3.1 \, .
\end{eqnarray}
The domination of the $WW$ channel above can be understood as the quintuplet contributes to both $H\to W^+ W^-$ and $ZZ$ channels while septuplet, as a real multiplet, contributes only to $H\to W^+ W^-$. The LHC 8 TeV run data constraints (\ref{run1}) are shown by grey area in Fig.~\ref{fig:param-ranges} where, like in the one-loop model case, $\gamma\gamma$ channel provides
the most stringent bound\footnote{We have also checked the 8 TeV data constraints from the remaining channels such as $t\bar{t}$ and di-jets.}.

\begin{figure}[h!]
\centerline{\includegraphics[scale=0.68]{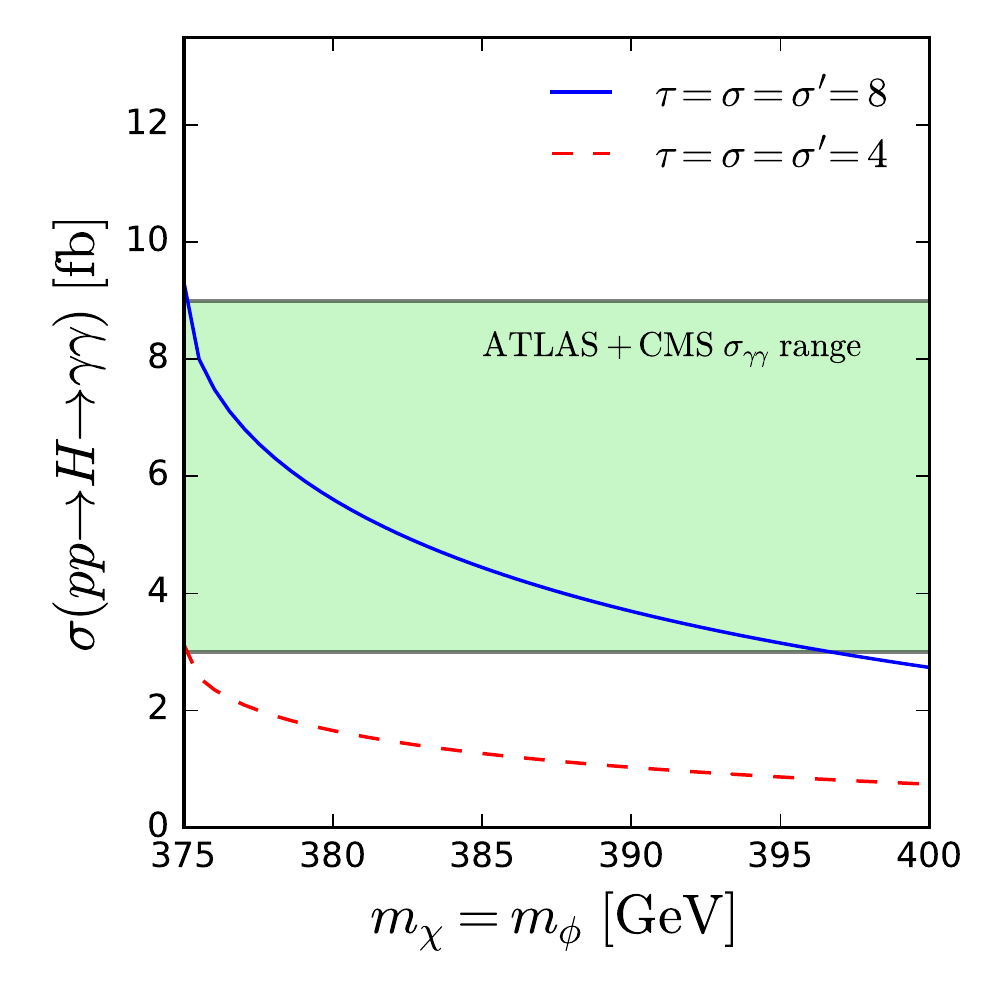}%
\includegraphics[scale=0.68]{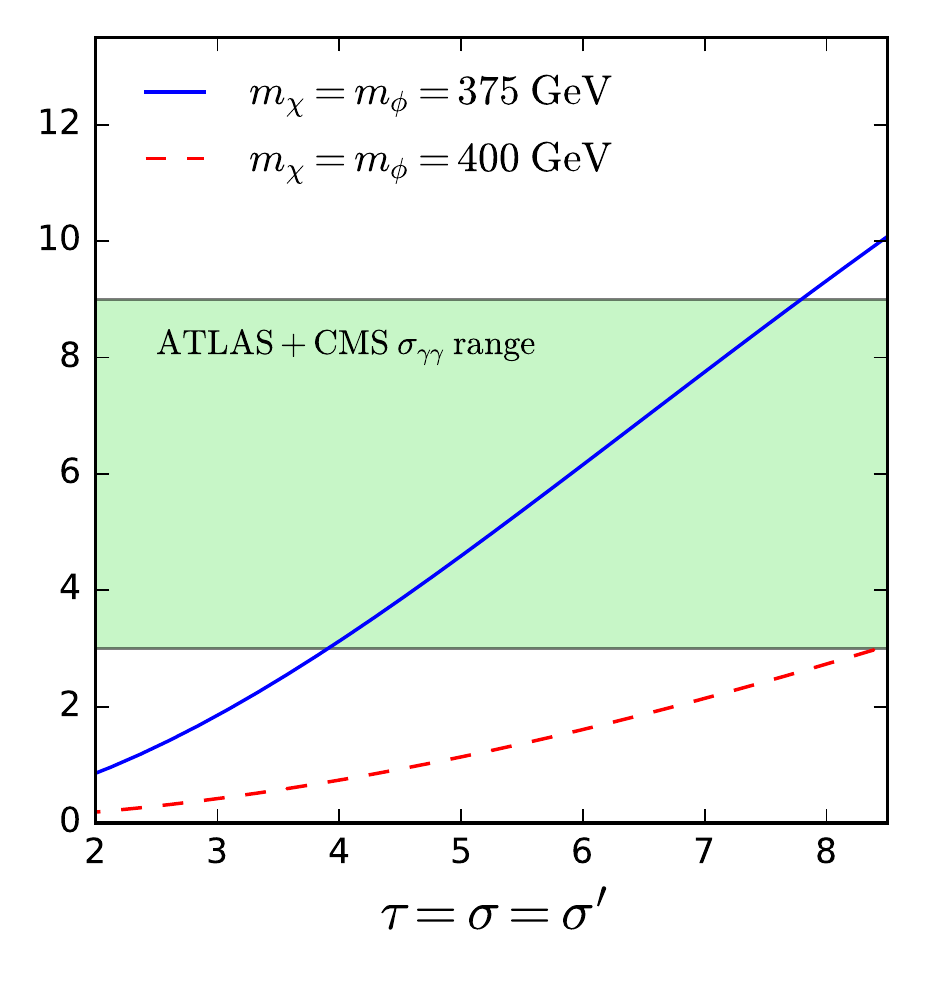}}
\caption{\label{photon75} Cross section for $p p \to H(750) \to \gamma\gamma$
in the three-loop neutrino mass model as a function of model parameters compared
to combined range of 3-9 fb for ATLAS and CMS diphoton anomaly (green).
}
\end{figure}

The total width of $H(750)$ for the generic choice of the parameters is shown in Fig.~\ref{fig:GAMTOT}. It is dominated by the $\bar{t}t$ channel, so that even in the extreme
case when $\tau=\sigma=\sigma'=8$, $m_\chi = m_\Phi = 375\,{\rm GeV}$, the branching ratio for diphoton channel is only
\begin{equation}
  Br(H\to\gamma\gamma) = 0.013 \,.
    \label{eq:BrggCKP}
\end{equation}
Intriguingly, in the range of parameter space where the model can accommodate diphoton cross-section, it also robustly predicts the large
total width of 30-50 GeV.
In Fig.~\ref{photon75} we show the 
diphoton cross section as a function of the parameters of the model compared to combined range of 3-9 fb for ATLAS and CMS diphoton anomaly.

\section{Vacuum stability and perturbativity}\label{perturbativity}

Minimal scenarios relying on extra singlet scalars and vector-like BSM fermions correspond to the particle content used in widely studied  class of ``simplified models for the Higgs physics" (e.g.~\cite{Dolan:2016eki}  and Refs. [21--29] therein).
By employing here a scalar field in the adjoint representation in the one-loop neutrino-mass scenario, we can only achieve the required diphoton signal strength for non-perturbative values of the couplings~\cite{Son:2015vfl} or for many copies of vector-like fermions. 
A summary of the detailed outcome of this model is presented in the first row in Table~\ref{Final}. We can contrast it to a recent claim~\cite{Badziak:2015zez}
that already one family of vector-like quarks and leptons with SM charges may be enough to explain the 750 GeV diphoton excess.

\begin{table}[h!]
\footnotesize
$$\begin{array}{lc|c|c|c||c|c|c|c|c|}
\hbox{Model}&\hbox{$J^{CP}_{750}$} & \hbox{$\Gamma_{750}$(GeV)}&\hbox{Production}  & \hbox{LP}  &\hbox{Br$_{WW}$}& \hbox{Br$_{\gamma\gamma}$}&\hbox{Br$_{Z\gamma}$}&\hbox{Br$_{ZZ}$}&\hbox{Br$_{t\bar{t}}$}\\
\hline
\rowcolor[cmyk]{0.1,0,0.1,0}
\multicolumn{1}{c}{\cellcolor[cmyk]{0,0,0.2,0} \hbox{1-loop}}  &0^{++} &\hbox{2.5-7.5} & \hbox{$\gamma\gamma$-fusion} &\hbox{Absent} &64\%\ & 7\%\ &6\%\ &23\%\ &- \\
\rowcolor[cmyk]{0.2,0.0,0,0.0}
\multicolumn{1}{c}{\cellcolor[cmyk]{0,0.2,0,0} \hbox{3-loop}} &0^{++}&\hbox{30-50}&\hbox{$gg$-fusion}& 10^6 \ \hbox{GeV} &23\%\ & 1\%\ &4\%\ &6\%\ &66\%\ \\
\end{array}
$$
\caption{ Comparison between the neutrino mass models. In the three-loop model the branching ratios are calculated for the benchmark point in (\ref{eq:BrggCKP}) leading to the total width $\Gamma_{750}\approx$ 45 GeV.}
\label{Final}
\end{table}

In the three-loop neutrino-mass scenario considered here, the charged components of exotic multicomponent scalar fields in a loop contribute to the diphoton decay of the neutral scalar in the 2HDM context,  as presented in the second row in Table~\ref{Final}.
We can contrast this to a recent three-loop radiative neutrino model with a local hidden U(1) symmetry~\cite{Ko:2016sxg} with another set of multiply charged  particles introduced to explain the 750 GeV diphoton excess.
\begin{figure}
\centerline{\includegraphics[scale=0.7]{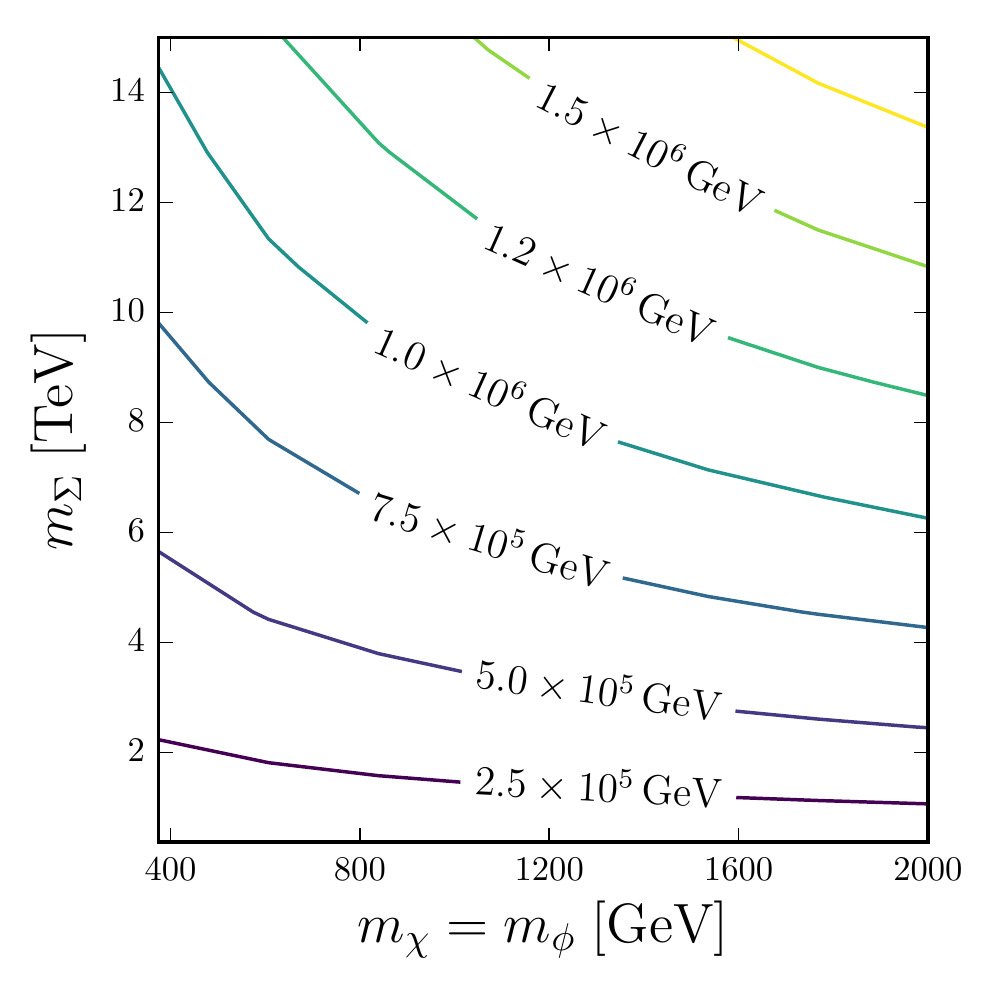}}
\caption{\label{fig:landau}\small Scale where the weak isospin coupling Landau pole appears in the three-loop neutrino
mass model in dependence of masses of new particles.
}
\end{figure}
The three-loop model at hand is under a well known threat that invoking large multiplets~\cite{Cirelli:2005uq} leads to Landau poles (LP) considerably below the Planck scale~\cite{Salvio:2016hnf}, potentially sensitive to two-loop RGE~\cite{DiLuzio:2015oha} effects.
For the $SU(2)_L$ gauge coupling $g_2$, this threat has been addressed in~\cite{Sierra:2016qfa} for the particle content of two scotogenic three-loop neutrino mass models~\cite{Culjak:2015qja,Ahriche:2015wha} aiming at accidental DM protecting $Z_2$ symmetry. Thereby the three-loop model at hand~\cite{Culjak:2015qja} is less affected by this threat, and its  exposure to additional scrutiny presented in  Fig.~\ref{fig:landau} shows that the LP appears around $10^6$ GeV.

As for the quartic couplings, the large values of the ``mixed" scalar couplings 
%of the $H_i^2 \Phi^2 (\chi^2)$ and $H_i^2 \chi^2$  types 
$\tau_{1,2}$ and $\sigma_{1,2}^{(\prime)}$ required to explain the di-photon excess and negative values for some of them from (\ref{alignment-constr-tau}) and (\ref{alignment-constr}), put the stability of the scalar potential and perturbative control over the model in danger. Here, we highlight the possible ways out of these difficulties.

First of all, we may depart from the limit of degenerate couplings, $\tau=\sigma=\sigma^\prime$, chosen for simplicity of the presentation in the previous section on the di-photon signal. 
In particular, we may choose initial value $\tau=0$ or $\sigma=\sigma^\prime=0$ at the particle threshold to turn off contributions to the di-photon signal from the septuplet $\chi$ or quintuplet $\Phi$, respectively\footnote{Of course, this will require even bigger contribution of the remaining ``mixed" quartic to the di-photon signal to compensate for the absence of the other multiplet.}. Related to this choice, we now discuss the different remedies that can be envisioned in the septuplet $\chi$ and the quintuplet $\Phi$ quartic sectors by activating them one at a time.

In the quartic sectors at hand, there are three additional quartic self-couplings of the $\Phi^4$-type and two additional quartic self-couplings of the $\chi^4$-type \cite{Hamada:2015bra} which we are still free to choose. There are additional quartics of the $\chi^2 \Phi^2$-type which we choose to be zero in order to decouple the septuplet and the quintuplet quartic sectors. 

Now, the stability of the potential will be endangered only due to those active ``mixed" quartics which are negative by the virtue of (\ref{alignment-constr-tau}) or (\ref{alignment-constr}), which may lead to an unbounded potential. Such quartics have to be balanced in the stability condition by appropriately chosen positive values of the corresponding quartic self-couplings (the stability condition for the septuplet sector has been explicated in \cite{Cai:2015kpa}). For the other inactive ``mixed" quartics we may choose the ``self" quartics to be zero at the threshold as well. 

As for the perturbative control of the model, it was shown in \cite{Hamada:2015bra} that for the inactive sector the LP will appear at:
\begin{eqnarray}
    \Lambda_{\Phi}\sim 10^9 \bigg(\frac{m_\Phi}{100 \  \hbox{GeV}} \bigg)^{1.28}\,{\rm GeV}\;, \qquad \Lambda_{\chi}\sim 10^6 \bigg(\frac{m_\chi}{100\  \hbox{GeV}}\bigg)^{1.13} \,{\rm GeV}\;,
\end{eqnarray}
for the quintuplet and the septuplet sectors, respectively. These values are not lower than $10^6$ GeV LP of the mentioned $SU(2)_L$ gauge coupling, so that we have a control over the inactive sector. As for the active scalar, we need to consider the possible Yukawa couplings of this scalar which provide a negative contribution to the one-loop beta function of the quartic self-couplings and may help to push the LP up. Unfortunately, for symmetry reasons, for the septuplet $\chi$ the obvious $\chi \Sigma\Sigma$ choice for the Yukawa term vanishes. Following \cite{Cai:2015kpa}, one may introduce the additional $SU(2)_L$-triplet fermion $\zeta=(3,0)$ to have a Yukawa coupling $\chi \Sigma\zeta$ which may be fine-tuned to delay the appearance of the LP. For the quintuplet $\Phi$, the needed Yukawa coupling $g_{i\alpha}$ already exists in our model in (\ref{yukawa}) and can be fine-tuned similarly.

Finally, the dominant contribution to the 1-loop beta functions of the ``mixed" quartics $\tau_{1,2}$ and $\sigma_{1,2}$ is given in \cite{Hamada:2015bra}:
\begin{equation}
\beta_{x} \sim 4 x^2 - \frac{153}{2} x g_2^2 + 36 g_2^4 \;, \qquad \beta_y \sim 4 y^2 - \frac{81}{2} y g_2^2 + 18 g_2^4 \;.
\end{equation}
Here, these couplings are denoted by $x=\tau_{1,2}$ and $y=\sigma_{1,2}$ and obey the conditions $\tau_1=-\tau_2$ and $\sigma_1=-\sigma_2$ from (\ref{alignment-constr-tau}) and (\ref{alignment-constr}). Due to large negative coefficients of $x g_2^2$ and $y g_2^2$ terms, it is easy to check that for $x<7.9$ and $y<4.1$ the sign of the beta function is such that by the running of the ``mixed" quartic coupling its initial value will be driven towards decreasing its absolute value\footnote{We took the SM value of the $SU(2)_L$ gauge coupling $g_2 (100 \ \hbox{GeV})\approx 0.65$. }.
As seen in Fig.\ref{photon75}, this parameter space overlaps with the values needed to explain the di-photon signal.
As we increase further the energy, the $SU(2)_L$ gauge coupling $g_2$ increases towards its LP and the $g_2^4$-term will eventually start to dominate the evolution, driving these ``mixed" quartics to the LP as well. We therefore expect that the dangerously-large initial values of the ``mixed" quartics needed to explain the di-photon signal will develop LP $\sim 10^6$ GeV together with the $g_2$ coupling.

\section{Discussion and conclusions}\label{Conclusions}

The very establishment of the SM is a successful bottom-up story: the Nature has been kind
to us in revealing the SM degrees of freedom, providing the answers to emerging questions gradually, one at a time. Additional BSM degrees of freedom seem to be most tangible when addressing the contemporary riddle of the  of neutrino-mass origin in the bottom-up way, since the BSM fields which  produce neutrino masses radiatively may be accessible at the LHC.\\
In the present account we take under scrutiny two radiative neutrino mass scenarios protected from tree-level contributions. An automatic $Z_2$ symmetry in the first (one-loop mass model) case forbids a tree-level mass contribution, and an accidental $Z_2$ symmetry in the second (three-loop mass model) case protects the stability of exotic BSM fields needed to close the three-loop mass diagram.\\

Additional arguments exposed in the previous section justify a hope that the three-loop mass model at hand may provide an appealing UV completion, in the same way as it is expected that the TeV-extensions of the SM would preserve the accidental baryon number of the SM to sufficient accuracy. 

Let us stress that the underlying $\tilde Z_2$ symmetry  imposed on the 2HDM potential~(\ref{2HDpot}) is exact as long as $m_{12}^ 2$,
$\lambda_6$ and $\lambda_7$ terms vanish. A detailed study  within the 2HDM scenario~\cite{Chakrabarty:2014aya,Dev:2014yca} shows that 
in the absence of the soft breaking $m_{12}^ 2$ term the exact $\tilde Z_2$ symmetry does not require intervention of new physics 
below $\sim$10 TeV scale. Indeed, at this scale the exotic states of three-loop scotogenic model~\cite{Culjak:2015qja} already enter into the play.
Despite the existence of the fortuitous DM-protecting symmetry $Z_2$, induced by $\tilde Z_2$ symmetry, the portion of the parameter space for the three-loop mass model which could reproduce the 750 GeV diphoton resonance seems to account only for a sub-dominant portion of the dark matter.

The hinted diphoton signal constrains the value of particular ``mixed'' quartic couplings of the model
as a welcome observable. On the other hand, a large value of this coupling leads to well known
Landau pole threat. Interestingly enough, there is another virtue of the aligned 2HD sector completed
with extra scalars in the context of the three-loop
model.
The mixture of the 2HD and exotic scalar sector provides a fortuitous remedy for the too early
Landau pole for relevant couplings, due to signs and sizes of the coefficients in the 
relevant beta functions.

To conclude, the existing hints of the diphoton resonance opened a hope that the history of
“prediscoveries” of new particles through the loop amplitudes may
be repeated in the scenarios taken under scrutiny. 
A verification of the diphoton signals at the LHC  may enable us to discriminate between scenarios offering different BSM fields.
On the other hand, if these hints
disappear with a larger integrated luminosity, they will still constrain the parameter space of proposed extensions of the SM with new charged states affecting considered loop amplitudes.

\vspace*{2ex}\noindent
\textbf{Note.} In the interest of open and reproducible research, computer
code used in production of final plots for this paper is made available at
\url{https://github.com/openhep/ackp16}.

\subsubsection*{Acknowledgment}
This work is supported by the Croatian Science Foundation under
the project number 8799, and by the QuantiXLie Center of Excellence.

\end{document}